\documentstyle[psfig]{l-aa}

\begin{document}
%\double
\thesaurus{3
	(11.06.2;
	11.12.2;
	13.18.1)}

\title{On the local radio luminosity function of galaxies.
I: the Virgo cluster}
\author{G. Gavazzi\inst{1}\and
A. Boselli\inst{2}}
\offprints{G. Gavazzi
	   Gavazzi @ trane.uni.mi.astro.it}
\institute{Universit\'a degli Studi di Milano, Via Celoria 16, 20133, Milano, Italy 
\and
Laboratoire d'Astronomie Spatiale, Traverse du Siphon, BP 8, F-13376 Marseille Cedex 12, France}

\date{Received , accepted }
\maketitle
%\markboth{The RLF of galaxies in Virgo}{...}

\begin{abstract}

We cross-correlate the galaxies brighter than $m_B=18$ in the Virgo cluster
with the radio sources in the NVSS survey (1.4 GHz), resulting in 180
radio-optical identifications. We determine the radio luminosity function
of the Virgo galaxies, separately for the early- and late-types. 
Late-type galaxies develop radio sources with a probability proportional
to their optical luminosity. In fact their radio/optical ($R_B$) distribution 
is gaussian, centered at $log R_B\sim -0.5$, i.e. the radio luminosity is 
$\sim$ 0.3 of the optical one. 
The probability of late-type galaxies to develop radio sources is almost
independent of their detailed Hubble type, except for Sa (and S0+S0a) which 
are a factor of $\sim$ 5 less frequent than later types at any $R_B$.\\
Giant elliptical galaxies feed "monster" radio sources with a probability strongly
increasing with mass. However the frequency of fainter radio 
sources is progressively less sensitive on the system mass. 
The faintest giant E galaxies ($M_B=-17$) have a probability
of feeding low power radio sources similar to that of dwarf E galaxies as faint as
$M_B=-13$. 
\footnote{Table 1 is only available in electronic form at
the CDS via anonymous ftp to cdsarc.u-strasbg.fr (130.79.128.5)
or via http://cdsweb.u-strasbg.fr/Abstract.html}


\end{abstract}

\section{Introduction}

A robust determination of the local 
(z=0) Radio Luminosity Function (RLF) of normal galaxies, jointly with 
similar determinations carried out at cosmological distances
(e.g. Prandoni et al. 1998),
is the essential tool for addressing several relevant cosmological issues,  
such as estimating the rate of evolution in galaxies,
more directly than by modeling the distribution of the faint radio
source counts (see Condon 1989).\\
Since the seventies this issue received strong attention among the 
scientific community, culminating with the analyses of elliptical
galaxies by Auriemma et al. (1977) and of spiral galaxies by Hummel (1981).
The study of the radio
properties of late-type galaxies in the Coma supercluster by Gavazzi 
\& Jaffe (1986) contributed 
establishing that the radio continuum luminosity of these galaxies is 
to first order
proportional to their optical luminosity, and to second order
to their current star formation rate. In other words, cosmic ray 
electron acceleration is provided primarily
by type II supernovae explosions (see also Condon 1992) which are more 
abundant in massive spiral galaxies. \\
Deep radio surveys, carried out in the nineties with the 
VLA, provided the mean
of re-determining the RLF of elliptical galaxies (Ledlow \& Owen 1996), 
which turned
out to be in remarkable agreement with the early determination of 
Auriemma et al. (1977).
Their main finding is that the probability of E galaxies to develop 
powerful radio sources ($log P_{1.4} > 24~W Hz^{-1}$) is strongly 
dependent on their optical luminosity ($L_B^{1.5}$). Below $log P_{1.4}=24~W Hz^{-1}$
this dependence becomes weaker with decreasing $P_{1.4}$.\\
Due to the lack of extensive radio and optical surveys of galaxies of both
early- and late-type, the present knowledge is still limited to relatively
bright radio sources ($log P_{1.4} > 21~W Hz^{-1}$) and bright optical 
luminosities ($M_B < -18.0$), typical of giant galaxies.\\
To go one step further
it seems natural to re-determine the RLF using the Virgo cluster, which 
contains thousands of galaxies, spanning a large luminosity range,
from giant to dwarfs as faint as $M_B=-13$ mag.
Surprisingly the latest systematic study of this cluster at centimetric 
wavelengths dates back 1981, when Kotanyi (1981) carried out with the WSRT
a survey of this cluster. Not only these early
measurement were limited by the current sensitivity (several mJy/beam), 
but also they lacked the necessary cluster coverage.\\
We are now in the position of re-addressing the issue 
taking advantage of the recent
all sky NVSS radio survey carried out with the VLA 
(Condon et al. 1998) and of the supreme quality of the Virgo Cluster 
Catalogue (VCC) of Binggeli et al. (1985), which provides us with reliable 
photometry and classification for over 2000 galaxies. 
The major improvement of the new radio observations is not primarily their higher sensitivity, which is in fact only few times better than 
previously available,
but mostly the unprecedented homogeneous sky coverage. \\
In this paper we make use of the NVSS data to construct the RLF of
an optically complete ($m_B < 18.0$) sample of galaxies extracted from 
the VCC. With these data we wish to discuss two issues:
i) is the dependence of the RLF on Hubble type well determined? 
ii) does the dependence of the RLF on galaxy mass, which is known to 
exist for giant galaxies, extend to the dwarf population?  
Issues i) and ii) are addressed in Section 4.\\
In a companion paper (Gavazzi \& Boselli
1999: Paper II) we address another question:
iii) is the local RLF of late-type galaxies universal or is it 
influenced by the environment? To study this issue we compare
the RLFs of late-type galaxies in five nearby clusters 
(Virgo, Cancer, A262, A1367 and Coma) with that of galaxies in less dense 
regions of the universe at similar distances.
$H_o=75~km~s^{-1} Mpc^{-1}$ is used throughout this paper.

\section{The Sample}

\subsection{The Optical Data}

The present investigation is based on the 
Virgo Cluster Catalogue (VCC) by Binggeli et al. (1985).
The VCC catalogue contains 2096 galaxies brighter than $m_B=20.0$. 
Photographic photometry with $\sim$ 0.35 mag uncertainty and
detailed morphological classification are given in the VCC.
The VCC coordinates are affected by $\sim$ 10 arcsec uncertainty (Binggeli et al.
1985) or slightly better.
The VCC also contains a (morphological) estimate of the membership 
to the various structures constituting the Virgo cluster:
cluster A (M87), cluster B (dominated by M49), W, W', M clouds, 
and Southern extension. General members and possible members are
treated in this work as belonging to the cluster. \\
We use an updated version of the VCC containing the following improvements:
i) for 305 galaxies listed in Binggeli \& Cameron (1993) we substitute the
eye estimated $m_p$ with $B_T$ obtained on digitized plates.
ii) we include the redshift and consequent membership re-assignments
given in Binggeli et al. (1993). We complement these data
with (few) other redshifts found in the New Extragalactic Database (NED).
iii) for 565 galaxies we substitute the celestial coordinates with the
more precise ones (few arcsec) listed in NED. By comparing this set of new
coordinates with the original VCC ones we find an rms difference of 
6.75 arcsec.
We extract from the catalogue a subsample of 1342 objects complete to 
$m_B=18.0$. Among these, 589 have yet no redshift in the literature 
(140 are possible members, 379 belong to cluster B and another 70 
are background galaxies).
Based on the most recent Cepheyds determination, we consider members of 
clusters A and B, general members, members of clouds W', M and Southern
extension at the distance of 17 Mpc. Members of W cloud are taken at
28 Mpc (see also Gavazzi et al. 1998). 

\subsection{1.4 GHz continuum data}

Radio continuum 1.4 GHz data in the regions covered by the present 
investigation are available from a variety of sources:\\
1) Full synthesis and snap-shot observations of specific regions 
were undertaken with the VLA and with the WSRT ("pointed" observations).
Hummel (1980) and Kotanyi (1980) did observations of the Virgo cluster 
with the WSRT. Condon (1987) and Condon et al. (1990) 
observed with the VLA nearby galaxies projected onto the Virgo 
region. These surveys do not generally constitute a complete set of 
observations.\\
2) Recently, the all-sky NVSS survey (Condon et al. 1998) carried out with 
the VLA at 1.4 GHz became available. 
The D array (FWHM = 45 arcsec) NVSS survey covers the sky north of 
$\delta >-40^{\circ}$,  with an average rms=0.45 mJy.
Except in specific regions
of the sky near bright sources, where the local rms is higher than average, 
this survey
offers an unprecedented homogeneous sky coverage. It not only provides us 
with extensive 
catalogues of faint radio sources, but also with homogeneous upper limits 
at any celestial position.   
The VCC region is covered by 13 NVSS maps which are available via
the World Wide Web.\\
Since radio data from more than one source exist for several target galaxies,
we choose between them adopting the following list of priorities:\\
1) in general we prefer NVSS data to any other source because of its 
homogeneous character,
relatively low flux density limit and because its FWHM beam 
better matches the 
apparent sizes of galaxies under study, thus providing us with flux 
estimates little affected by missing extended flux.\\
2) For individual bright radio galaxies (e.g. M87) we prefer
data from specific "pointed" observations since they should provide us 
with more reliable estimates of their total flux.\\
3) in all cases where the flux densities from NVSS are lower than those 
given in other references we privilege the reference carrying the 
largest flux density.

%Fig 1
\begin{figure}
\vbox{\null\vskip 13.0cm
\includegraphics{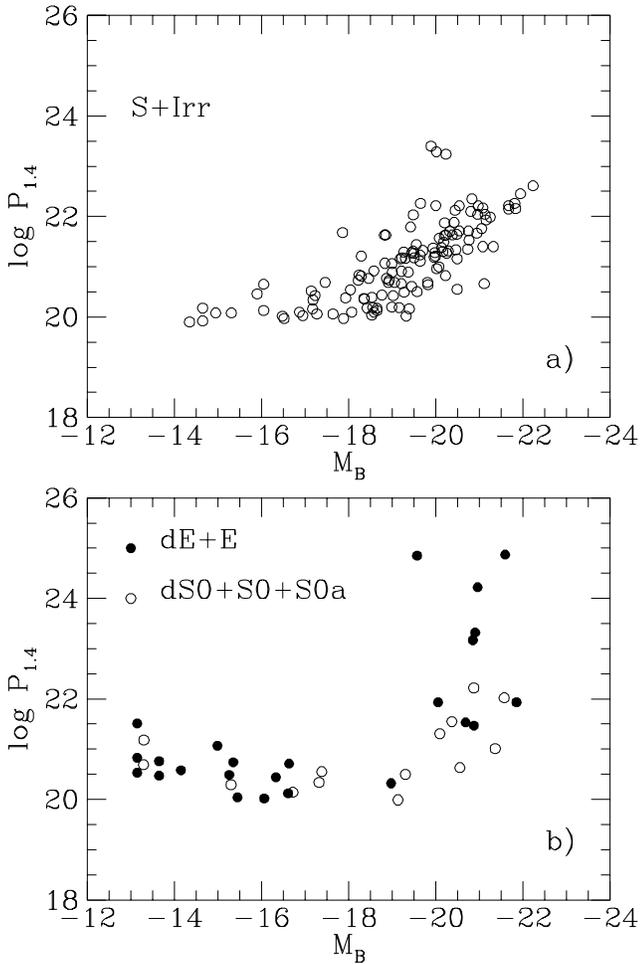}
}
\caption{The correlation betwen the radio and optical luminosity for the detected
galaxies. a: late-type; b: early-type.
} 
\label{Fig.1}
\end{figure}

\subsection{The radio-optical identifications}

At the position of all optically selected galaxies  
we search for
a radio-optical coincidence. For the remaining undetected galaxies 
we compute an upper limit flux using $4\times rms$.
For the purpose of our study we proceed as follows:\\
1) we pre-select sources from the NVSS data-base, allowing for a maximum 
radio-optical positional 
discrepancy of 45 arcsec, because of the large apparent size of the galaxies 
in this nearby cluster.\\
2) we inspect the NVSS maps at the position
of all candidate radio-optical associations. We discard few spurious radio 
sources listed in the NVSS database, which turn out to be grating rings 
associated with strong radio sources. These belong to 3 regions of
approximately 40 arcmin radius around M87, NGC4261 (3C-270), and around 
the source $12^h30^m+02$. In these regions we also re-compute the local 
rms radio noise.\\
3) at the position of all pre-selected optical-radio matches we compute an 
"identification
class" (ID) according to a criterion which is a slight modification of the 
one adopted by Jaffe \& Gavazzi (1986). 
For each galaxy we calculate the quantity:\\
\noindent
 $R^2={\Delta_{r-o}^2\over\sigma_g^2+\sigma_{r}^2}$~~~~~~~~~~~~~~~(1)\\
where $\Delta_{r-o}$ is the radio - optical positional offset, 
$\sigma_{r}$ is the radio position uncertainty,
and $\sigma_g$ is the uncertainty in the galaxy position.
The latter quantity is assumed to be the 3\% of the galaxy optical diameter 
plus the
uncertainty in the optical position itself:\\
\noindent
 $\sigma_g^2=0.0009\times A^2+\sigma_o^2$~~~~~~~~~~~~~~~(2)\\
\noindent
where A is the galaxy optical major axis, and $\sigma_o$ is 
the uncertainty in the optical position (see Section 2.1).
\noindent 
The errors on the radio positions are assumed to be inversely proportional 
to the signal-to-noise ratio as:\\
\noindent
 $\sigma_{r}=0.5 \times FWHM \times \sqrt{rms/flux}+0.3$~~~~~~~~~(3)\\
\noindent
Identification class ID=1 includes pointlike radio sources with $R\leq3$.\\
\noindent
Identification class ID=2 are extended sources not meeting the $3\sigma$ 
criterion.\\
\noindent
Identification class ID=3 are dubious identifications not meeting the 
$3\sigma$ criterion (not used in the following analysis).\\
\noindent
Identification class ID=4 include pointlike sources whose radio-optical 
offset is within
the optical extent of the galaxy. These are dubious identifications, 
which are nevertheless used
because off-set radio sources are often found associated with disk galaxies. \\
\noindent

%Fig 2
\begin{figure}
\vbox{\null\vskip 7.8cm
\includegraphics{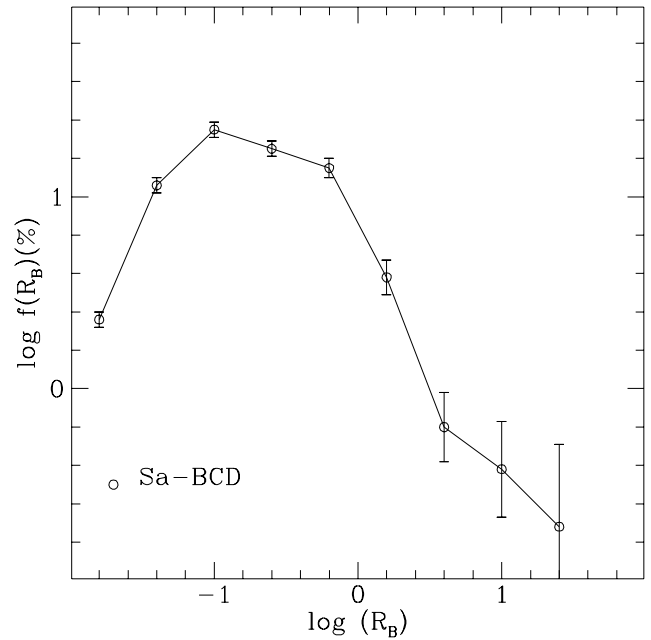}
}
\caption{the differential RLF as a function of the radio/optical ratio $R_B$ for all
late-type galaxies (Sa-BCD).
} 
\label{Fig.2}
\end{figure}

%Fig 3
\begin{figure*}
\vbox{\null\vskip 16.0cm
\includegraphics{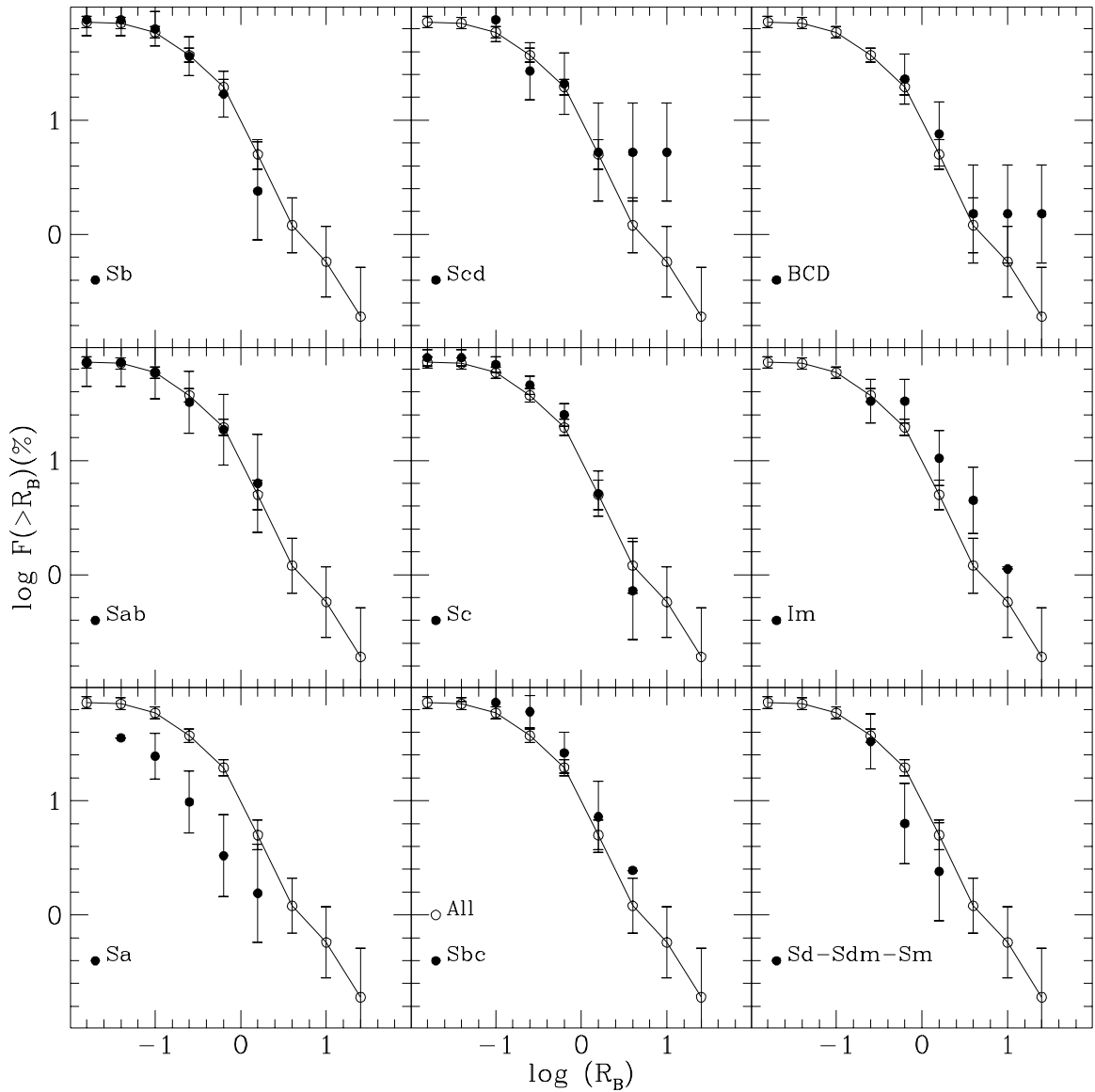}
}
\caption{the cumulative RLF as a function of the radio/optical ratio $R_B$ of
late-type galaxies is given separately for each morphology class (filled dots). The distribution
of all late-type is also given for comparison (open dots).
} 
\label{Fig.3}
\end{figure*}

The 180 positive radio-optical matches are listed in Table 1 as follows: \newline
Column 1: the VCC (Binggeli et al. 1985) designation.\newline
Column 2: the photographic magnitude corrected for extinction in
our Galaxy according to Burstein \& Heiles (1982) and for internal extinction
following the prescriptions of Gavazzi \& Boselli (1996), except that the correction
for internal extinction has been omitted for Irr and dwarfs.\\
Column 3: the morphological classification as given in the VCC.\newline
Column 4: the membership to the individual clusters and clouds in the 
Virgo area as given in Binggeli et al. 1985 and revised in Binggeli et al. (1993).\\
Columns 5, 6: the (B1950) optical celestial coordinates of the target galaxy.\newline
Columns 7, 8: the (B1950) celestial coordinates of the radio source.\newline
Column 9: the radio-optical offset (arcsec).\newline
Columns 10: the identification class (see above).\\
Column 11: the 1.4 GHz total flux density (mJy).\\
Columns 12, 13: the extension parameters of the radio source (major and minor axes
in arcsec).\\
Column 14: reference to the 1.4 GHz data. All except 12 identifications
are based on NVSS data.\\

%Fig 4
\begin{figure*}
\vbox{\null\vskip 6.0cm
\includegraphics{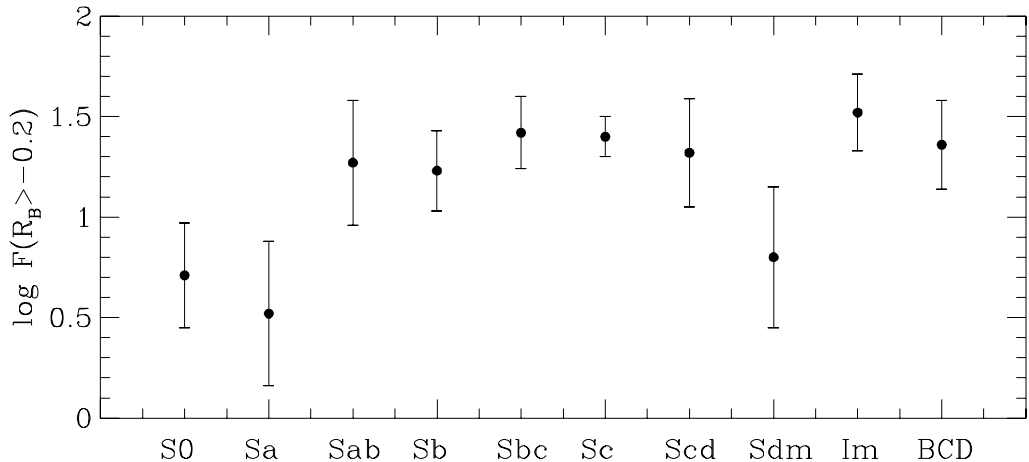}
}
\caption{the cumulative RLF for $log R_B > -0.2$ is given as a function of the Hubble type.
} 
\label{Fig.4}
\end{figure*}

All our ID=1 sources are found within 35 arcsec from the central optical
coordinates of the parent galaxies. Some (generally fainter than 10 mJy) 
ID=2 and ID=4 sources lie between 35 and 45 arcsec. 
An estimate of the number of possible chance-identifications ($N_{c.i.}$) 
among the 180 sources/galaxies listed in Table 1 is carried out with two 
independent
methods. Using Condon et al. (1998) Fig. 6 we estimate that the probability 
of finding an unrelated source within 45 arcsec of an arbitrary position
is 2~\%. Thus about 3.6 sources in Table 1 should be spurious associations.
An independent estimate is computed according to:\\
 $N_{c.i.}=9\pi\times \Sigma\sigma_r^2\times N_g/A $~~~~~~~~~~~~~~~(4)\\
where the summation is extended to all radio sources (approximately 7000)
found in the area A ($140 deg^2$) containing the Virgo cluster, and
$N_g$ is the total number of galaxies considered (1342). $N_{c.i.}$
is 5.3, in good agreement with the previous determination.

%Fig 5
\begin{figure*}
\vbox{\null\vskip 11.5cm
\includegraphics{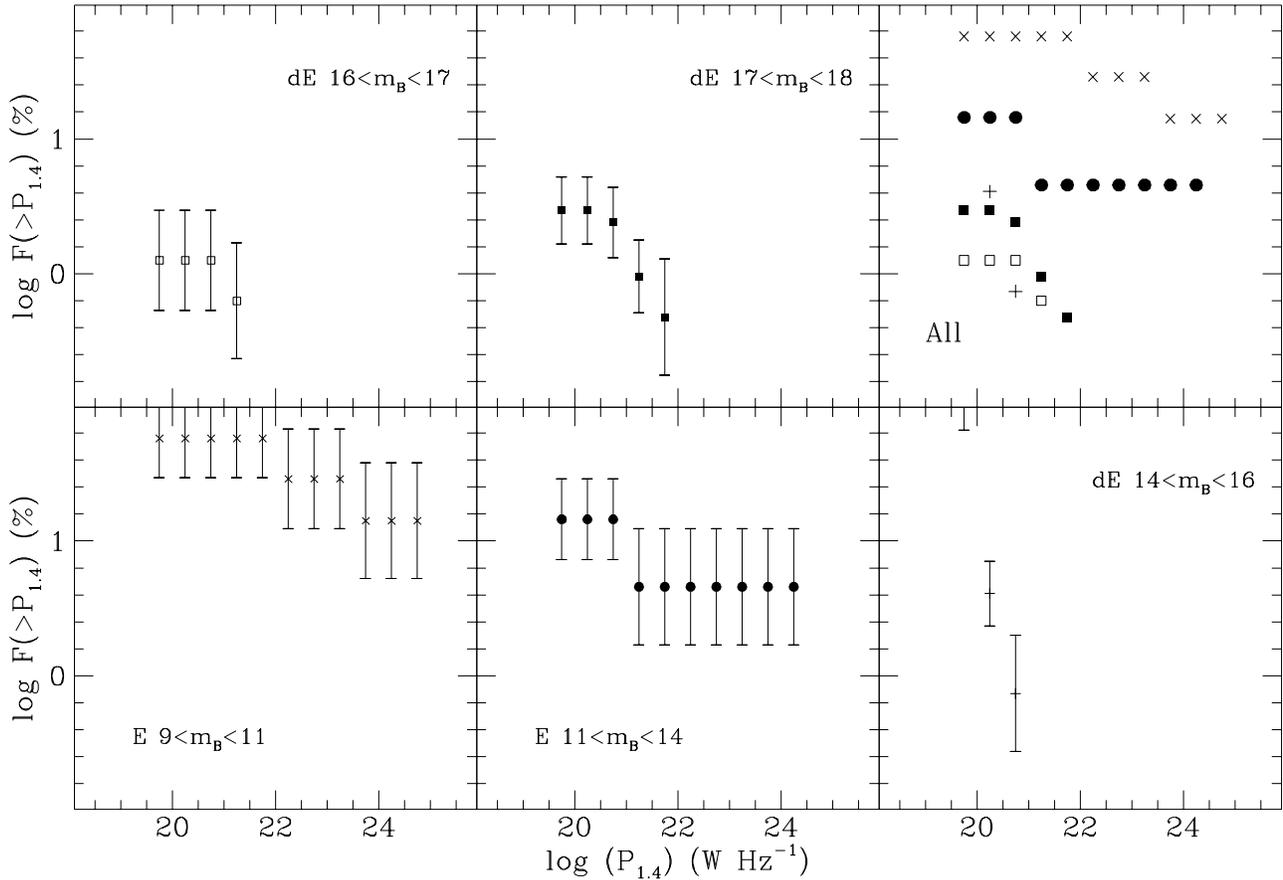}
}
\caption{the cumulative RLF of Elliptical galaxies in five bins of optical luminosity.
} 
\label{Fig.5}
\end{figure*}

\section{The radio-optical luminosity}

For all galaxies we have a measurement of the total 
radio flux density 
$S_{1.4}$ (for 180 detections) or an upper limit $S_l$ (undetected objects). 
Combining this with the distance to the objects
we determine their radio luminosity $P_{1.4}$ (or $P_l$) in $W Hz^{-1}$.\\
Let us consider the detected galaxies first. The existence of a 
correlation between the radio and optical luminosity is well known:
(see e.g. Condon 1980; Gavazzi \& Jaffe
1986; Gavazzi \& Contursi 1994). Figs. 1 a and b show this correlation separately for the
early and late-types. 
In spite of a large scatter which indicates that,
beside the optical magnitude other quantities determine the radio
properties of galaxies, it appears that late-type galaxies obey to an
almost direct proportionality between the two quantities.
Early type galaxies instead show a strongly non-linear behavior:
galaxies fainter than approximately $M_B=-19$ have an average radio luminosity 
$\sim 10^{21}~W Hz^{-1}$, independent of their optical luminosity. Galaxies brighter
than $M_B=-19$ have a radio luminosity strongly increasing with $M_B$.
These are the "monster" radio galaxies. It is interesting to notice in Fig.
1b that "monsters" are absent among S0+S0a galaxies.

\section{The Radio Luminosity Functions}

Given the small number of actual radio detections (180) with respect to the number
of optical candidates (1342), Fig. 1 might not give a realistic representation
of the radio properties of an optically selected sample of galaxies. More appropriately, that can
be derived in the form of the Fractional Radio Luminosity Function (RLF), which gives the 
probability distribution 
$f(P)$ that galaxies develop a radio source of a given luminosity ($P_{1.4}$),
taking into 
account the number of detected objects $n_d(P_k)$ in each bin of 
radio 
luminosity $P_k$, as well as the upper limits (see Avni et al. 1980).

The differential distribution can be derived adopting method III of Hummel 
(1981) 
which has been shown to be equivalent to the expression given by 
Avni et al. (1980):\\
\noindent
 $f(P_k)={n_d(P_k)\times (1- \sum_{j=1}^{k-1} f(P_j)) \over
n_u(P_l<P_k)+n_d(P\leq P_k)}$~~~~~~~~~~~~~~~(5)\\
\noindent
where:\\
\noindent
$n_d(P\leq P_k)$ is the number of detected objects with $P\leq P_k$,
$n_u(P_l<P_k)$ is the number of undetected objects with $P_l<P_k$.\\
\noindent
The uncertainty on $f(P_k)$ is given by:\\
\noindent
 $\sigma f(P_k)~=~{f(P_k)\over \sqrt{n_d(P\geq P_k)}}$~~~~~~~~~~~~~~~(6)\\
\noindent
The cumulative distribution is thus:\\
\noindent
 $F(\geq P_k)= \sum_{j=1}^{k} f(P_j)$~~~~~~~~~~~~~~~(7)\\

\subsection{Late-type galaxies}

Spiral galaxies are well known to develop radio sources with an average 
radio luminosity
proportional to their optical luminosity (see Section 3). For these objects it is 
convenient
to define the (distance independent) radio/optical ratio: 
$R_B= S_{1.4} / k\times 10^{-0.4*m_B}$,
where $m_B$ is the B magnitude and $k=4.44\times 10^6$ is the factor appropriate
to transform the broad-band B magnitudes in mJy.
$R_B$ gives the ratio of the radio emission per unit light emitted 
by the relatively young stellar population.
The distribution of $f(R_B)$, that is the probability that a 
galaxy develops a radio source with
a radio/optical ratio $R_B$ can be derived similarly to $f(P_{1.4})$ 
using equation (5).\\
Since the radio/optical ratio is a distance independent quantity, we extend
the present analysis to all galaxies in the VCC, including the background
objects. The exclusion of these objects reduces significantly the statistics,
without changing the results.\\
The $f(R_B)$ for all galaxies (from Sa to BCD), given in Fig 2, 
appears as a normal distribution peaked at Log $f(R_B)$ between -1 and -0.5.
About 20 \%
of all galaxies are detected at the peak of the distribution. 
At the sensitivity of
the present radio survey about 70 \% of Virgo galaxies have $R_B>0.01$.
Let us now consider separately the following 9 morphological type classes:
Sa, Sab, Sb, Sbc, Sc, Scd, Sd+Sdm+Sm, Im, and BCD.
Barred and ringed spirals are mixed with normal spirals.\\
The cumulative $F(\geq R_B)$ distributions 
are shown in 9 panels of Fig. 3 respectively.
All panels show, for comparison, also the distribution of all late-type 
galaxies together (open dots connected with a solid line).
It is striking that, within the statistical uncertainties,
all RLFs are consistent among each other, except for the Sa's. 
The Sa's develop radio sources, at any given value of $R_B$, about 5 times 
less
frequently than all others late-type galaxies. Sd+Sdm+Sm are slightly underluminous than average, 
but this difference is barely significant.
We have also computed the $F(\geq R_B)$ distribution of S0+S0a (not shown
in Fig. 3), which results identical to that of Sa galaxies. \\
Fig. 4 summarizes the dependence of the radio properties on Hubble type
using the cumulative fraction $F(> R_B=-0.2)$.
These results are in full agreement with the findings of Hummel (1981).\\

%Fig 6
\begin{figure}
\vbox{\null\vskip 8.0cm
\includegraphics{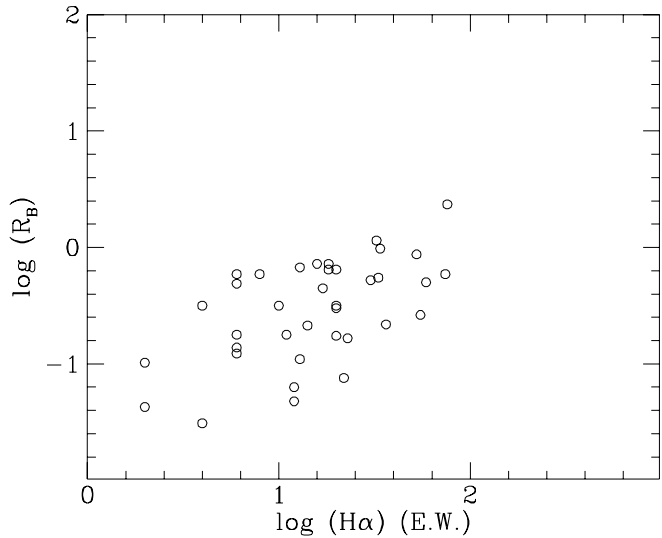}
}
\caption{the relation between the radio/optical ratio $R_B$ and
the equivalent width of the $H_\alpha$ line for the late-type detected galaxies.
} 
\label{Fig.6}
\end{figure}

\subsection{Early-type galaxies}

Early-type galaxies do not develop radio sources with a radio luminosity
proportional to their optical luminosity (see Section 3). For these objects
the radio/optical ratio is meaningless.
This is consequent to the very existence of radio galaxies. These galaxies 
develop radio sources with a broad range of power, whose nature is nuclear, 
thus it is only indirectly related with the total luminosity 
of their host galaxies.
For early-type galaxies it is convenient to analyze the "bivariate" (i.e. per
interval of optical luminosity) distribution of 
radio luminosity. Obviously this analysis is restricted to the
members of Virgo, disregarding the background objects.\\
The cumulative representation $F(> P_{1.4})$ is shown in Fig. 5 in
5 bins of optical luminosity, such that each of them contains a significant
number of objects: $9<m_B<11$ (7); $11<m_B<14$ (22); $14<m_B<16$ (141);
$16<m_B<17$ (159) and $17<m_B<18$ (211).
Within the range of radio power covered by the present
analysis ($20 < log P_{1.4} < 25~WHz^{-1}$) it appears that the probability for E galaxies
to develop radio sources decreases steeply with the optical luminosity
only above $m_B=16$ (which corresponds to $M_B=-15$). 
Below this optical luminosity,
where dEs dominate, the fraction $F(> P_{1.4})$ becomes independent of 
the optical luminosity.

%Fig 7
\begin{figure*}
\vbox{\null\vskip 10.0cm
\includegraphics{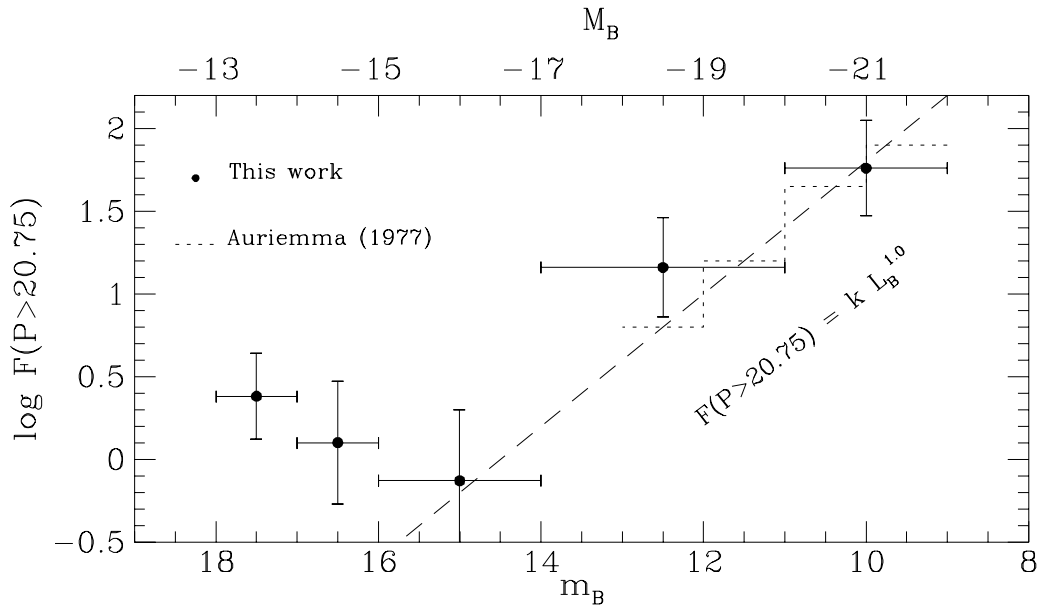}
}
\caption{the cumulative fraction of elliptical galaxies with radio luminosity $log P_{1.4}>20.75$
as a function of optical luminosity. This work (filled dots); Auriemma et al. (1977) (dotted
line).
} 
\label{Fig.7}
\end{figure*}

\section{Discussion}

A clearcut result of the present investigation is that spiral-Irr galaxies
develop extended radio sources whose luminosity scales with the optical luminosity,
independently of the detailed Hubble type.
Their radio/optical ratio is linearly correlated with
their current, massive star formation rate, as derived from their
$H_\alpha$ emission line intensity (Kennicutt, 1999). 
Fig. 6 shows that, among the detected galaxies, the relation between $R_B$ and
the equivalent width of the $H_\alpha$ line is one of direct
proportionality, implying that cosmic-ray acceleration is primarely associated with
type II supernovae.\\
Elliptical galaxies develop radio sources whose nature is nuclear.
Auriemma et al. (1977) determined the RLF of E galaxies brighter of $M_B=-18$. 
They found that the RLF shows a
pronounced break at the radio luminosity $log P_{1.4}^* = 24~WHz^{-1}$, independent of the 
galaxy optical luminosity. Beyond the break $F(> P_{1.4})$ scales
with the optical luminosity as $L_B^{1.5}$. Below $P_{1.4}^*$, instead
the dependence of $F(> P_{1.4})$ on $L_B$ is weaker.
The RLF determined in this work extends to optical luminosities 5
magnitude fainter than Auriemma et al. (1977). However, it does
not comprise radio luminosities greater than $log P_{1.4} = 24~WHz^{-1}$. The
only powerful radio galaxy in Virgo is M87, which is in fact
right above $P_{1.4}^*$.
With our data we can study how the radio properties of E galaxies
scale with the optical luminosity below $P_{1.4}^*$.
The dependence of the probability for E galaxies
to develop radio sources with $log P_{1.4}>20.75~WHz^{-1}$ is represented in Fig. 7
as a function of the optical luminosity. 
Together with our data, Fig. 7 also represents the
results of Auriemma et al. (1977) (dotted line, adapted from their 
Fig. 5) which appear in full agreement with ours.
The three brightest points ($9<m_B<16$) are
well represented by $log P_{1.4} \sim L_B^{1.0}$; however, below $m_B=16$, the data 
are consistent with no further dependence from $L_B$. 
It is instructive to compare Fig. 1b with Fig. 7, which contain complementary
information: Fig. 1 shows the dependence of the radio on the optical luminosity
of the detected objects, while Fig. 7 adds the information on the frequency
at which Early-type galaxies develop radio sources (with a given radio luminosity)
as a function of the optical luminosity. 
We conclude that, while the frequency with which E galaxies feed
the "monsters" in their nuclei is strongly related to their total
mass, that of fainter radio sources is progressively less sensitive on the
system mass. In fact it does not decrease from $M_B=-16$ to 
$M_B=-13$. The faintest giant E galaxies have similar probability
of feeding low power radio sources than dwarf E galaxies 3-4
mag fainter. 

\section{Summary}

In summary the present investigation brought us to the following empirical results:\\
1) Late-type galaxies develop radio sources with a probability proportional
to their optical luminosity. In fact their radio/optical ratio is a gaussian
distribution centered at $R_B\sim -0.5$, i.e. the radio luminosity is $\sim$ 0.3
of the optical one. About 20 \% of all spiral galaxies is detected at the peak
of the distribution.\\
2) The probability of late-type galaxies to develop radio sources is almost
independent of their detailed Hubble type, except that Sa (and S0+S0a) are at
least a factor of $\sim$ 5 less frequent at any value of radio/optical.\\
3) The relation between $R_B$ and the equivalent width of the $H_\alpha$ line 
is of direct proportionality.\\
4) The luminosity of radio sources associated with Early-type galaxies increases
non-linearly with the optical luminosity of their parent galaxies.\\
5) The probability of finding low luminosity ($log P_{1.4}>20.75~WHz^{-1}$) radio sources
associated with Early-type galaxies scales non-linearly with the optical luminosity.

\acknowledgements {We wish to thank P. Pedotti for her contribution 
to this work, T. Maccacaro for useful discussions and 
B. Binggeli for providing us with the VCC in digital form. 
This work could not be completed without access to the New Extragalactic Data-
Base (NED) which is operated by the Jet Propulsion Laboratory, California Institute
of Technology, under contract with the National Aeronautics and Space Administration. 
We wish also to acknowledge J. Condon and the NVSS team for their
magnificent work.}


\begin{thebibliography}{}  
\bibitem[]{}
Auriemma C., Perola C., Ekers R., et al. 1977, A\&A, 57, 41
\bibitem[]{}
Avni Y., Soltan A., Tananbaum H., Zamorani C., 1980, ApJ, 238, 800
\bibitem[]{}
Binggeli B., Sandage A., Tammann G., 1985, AJ, 90, 1681 (VCC)
\bibitem[]{}
Binggeli B., Popescu C., Tammann G., 1993, A\&AS, 98, 275
\bibitem[]{}
Binggeli B., Cameron L., 1993, A\&AS, 89, 297
\bibitem[]{}
Burstein D., Heiles C., 1982, AJ, 87, 1165
\bibitem[]{}
Condon J., 1980, ApJ, 242, 894
\bibitem[]{}
Condon J., 1987, ApJS, 65, 485
\bibitem[]{}
Condon J., 1989, ApJ, 338, 13
\bibitem[]{}
Condon J., Helou G., Sanders D., Soifer B., 1990, ApJS, 73, 359
\bibitem[]{}
Condon J., 1992, ARA\&A, 30, 575
\bibitem[]{}
Condon J., Cotton W., Greisen E., et al., 1998, AJ, 115, 1693 (NVSS)
\bibitem[1986]{Gav}
Gavazzi G., Jaffe W., 1986, ApJ, 310, 53
\bibitem[]{}
Gavazzi G., Contursi A., 1994, AJ, 108, 24
\bibitem[1996a]{GBcat}
Gavazzi G.,  Boselli A., 1996, Astroph. Lett \& Commun, 35, 1
%\bibitem[1996b]{GPB96}
%Gavazzi G., Pierini D., Boselli A., 1996, A\&A, 312 397
\bibitem[]{}
Gavazzi G., Boselli A., Scodeggio M., Pierini D., Belsole E., 1998, MNRAS (in press)
\bibitem[]{}
Gavazzi G., Boselli A., 1998, A\&A (in press) (Paper II)
\bibitem[]{}
Jaffe W., Gavazzi G., 1986, AJ, 91, 204
\bibitem[]{}
Hummel E., 1980, A\&AS, 41, 151
\bibitem[]{}
Hummel E., 1981, A\&A, 93, 93 
\bibitem[]{}
Kennicutt R., 1999, ARA\&A, (astro-ph/9807187)
\bibitem[]{}
Kotanyi C., 1980, AAS, 41, 421
\bibitem[]{}
Ledlow M., Owen F., 1996, AJ, 112, 9 
\bibitem[]{}
Prandoni I., Gregorini L., Parma P., et al., 1998, in {\it Looking Deep in the Southern Sky},
eds. Morganti \& Couch (Springer-Verlag), in press. 

\end{thebibliography}
\end{document}